\documentstyle[preprint,aps,prb,epsfig,amsmath,amssymb]{revtex}

\begin{document}

\draft

\title{Crossover Scales at the Critical Points of Fluids with
Electrostatic Interactions}

\author{Andr{\'e} G. Moreira \footnote{Current address:
Max-Planck-Institut 
f{\"u}r Kolloid- und  Grenzfl{\"a}chenforschung, Kantstr. 55, 14513 Teltow, 
Germany.}  \and M. M. Telo da Gama}

\address{Departamento de F{\'\i}sica da Faculdade de Ci{\^e}ncias and 
Centro de F{\'\i}sica da Mat{\'e}ria Condensada\\
Universidade de Lisboa, 
Avenida Professor Gama Pinto 2, P-1649-003 Lisboa, Portugal}

\author{Michael E. Fisher}

\address{Institute for Physical Science and Technology,\\University of
Maryland, College Park, Maryland 20742, U.S.A.}

\maketitle

\begin{abstract}

Criticality in a fluid of dielectric constant $D$ that exhibits Ising-type
behavior is studied as additional electrostatic (i.e., ionic) interactions
are turned on. An exploratory perturbative calculation is performed for
small {\em ionicity} as measured by the ratio of the electrostatic
energy
$e^2/D \, a$ (of two univalent charges, $\pm e$, separated by the
atomic/ionic
diameter $a$) to $k_B \, T_c^0$ which represents the strength of the
short-range {\em non}ionic (i.e., van der Waals) interactions in the
uncharged fluid. With the aid of distinct transformations for the
short-range and for the Coulombic interactions, an effective Hamiltonian
with coefficients depending
on the ionicity is derived at the Debye-H{\"u}ckel limiting-law level
for a fully symmetric model. The crossover between classical (mean-field)
and Ising behavior is then estimated using a Ginzburg criterion.
This indicates that the reduced crossover temperature depends only weakly
on the ionicity (and on the range of the nonionic potentials); however,
the trends do correlate with the, much stronger, dependence observed
experimentally. 
\end{abstract}



\section{Introduction}


Early studies of the liquid-vapor critical point of ionic systems were
carried out on molten salts and related systems,   
\cite{Kirshenbaum-al:62,Chieux-Sienko:70,Buback-Franck:72} with
relatively poor temperature stability and resolution. 
\cite{LeveltSengers-Given:93}
Results for ionic critical behavior at low reduced temperatures were first 
published by Pitzer and co-workers a decade ago. 
\cite{Pitzer-deLima-Schreiber:85}
Subsequently other solutions of organic salts, with low critical
temperatures, were studied yielding somewhat puzzling results.

One of the systems studied in detail was triethyl-$n$-hexylammonium 
triethyl-$n$-hexylboride (N$_{2226}$--B$_{2226}$) in diphenyl ether. In 
1990, Singh and Pitzer\cite{Singh-Pitzer:90} reported a coexistence-curve 
exponent, $\beta \simeq 0.476$, close to the mean field value of
$\frac{1}{2}$. Their data are compatible with Ising criticality ($\beta
\simeq \frac{1}{3}$) only if the 
mean field to Ising crossover occurs at a reduced temperature, $t_\times
\equiv
|T-T_c|/T_c \simeq 10^{-4}$. \cite{Singh-Pitzer:90,Fisher:94} This
value
of $t_\times$ is unusually
small---in ordinary
fluids $t_\times \sim 1$ and the observed critical exponents are clearly  
of Ising character.
In 1992, Zhang {\it et
al.} \cite{Zhang-al:92} confirmed the mean-field critical behavior of this
system
by reporting a susceptibility exponent, $\gamma \simeq 1.0$. \cite{Wiegand-al:97} 

A different system also studied nearly a decade ago was
tetra-$n$-pentylammonium
bromide dissolved in water. Japas and Levelt Sengers
\cite{Japas-LeveltSengers:90}
reported an Ising-type  coexistence curve  with $\beta=0.319$--$0.337$. 
Other studies reported Ising or mean-field
criticality depending on the particular ionic solution. For 
reviews of the various experiments, see: Refs.  
\cite{LeveltSengers-Given:93,Fisher:94,Pitzer:95,Weingartner_1:95}

The presence of the solvent, which lowers the critical temperature of the
salts, complicates
the situation considerably. The bare Coulomb interaction is reduced by a
factor that depends on, $D$, the static dielectric constant of the
solvent. Additionally, in polar solvents, solvation shells are formed
around each ion and the resulting effective interactions between the ions
are not well characterized at present.
The mechanisms driving criticality in these various systems are not 
clear, but there appear to be two classes of
critical behaviour: Coulombic systems, where the electrostatic
interactions primarily drive the phase transition and solvophobic
mixtures in which the phase transition is mainly driven by the 
short-ranged
interactions that favor solvent-solvent and solute-solute contacts. 
\cite{LeveltSengers-Given:93,Fisher:94,Pitzer:90,Weingartner:91}

In Table \ref{tab:Tm_Tc} we display data on the melting temperature of
three salts and their critical temperatures in solution. Clearly, if $T_m$
is \emph{lower} than $T_c$ the Coulomb interactions are too weak
to drive the phase transition, especially in high dielectric-constant
solvents. \cite{Japas-LeveltSengers:90}
However, there are cases where the ionic interactions are strong enough
to play a role at temperatures of the order
of $T_c$.  In these systems,  
\cite{Narayanan-Pitzer:94} 
a crossover from mean-field to Ising criticality was observed, with
reduced
crossover temperatures $t_\times$, that decrease as the relative ionic
character or {\em ionicity} increases, i.e., the dielectric
constant of the solvent decreases. Ising exponents are observed only 
closer
to $T_c$ as the ionicity increases.
It is plausible to suppose that the effect of the
Coulomb interactions is the reduction of the crossover
temperature; this reduction might be so strong that Ising
critical behavior could not be observed in practice. Ideal Coulombic
electrolyte
criticality
would then be
characterized by classical (mean-field) behavior, while solvophobic
electrolyte 
criticality remained in the 3D Ising universality class.
\cite{Pitzer:90,Weingartner:91}

However, a theoretical description of the dependence of $t_\times$ on the
Coulombic and other interactions is still lacking. Indeed, previous
attempts to study the
crossover temperature of model ionic fluids do not seem to predict any
reduction, at least for the restrictive primitive model (RPM) consisting of an
equal number of equisized positive and negative hard-spheres. Following
the sugestion of Fisher and Levin, \cite{Fisher-Levin:93} 
Leote de Carvalho and Evans \cite{Carvalho-Evans:95} and Fisher and 
Lee \cite{Fisher-Lee:96} employed a Ginzburg criterion which led them to
conclude that the reduced crossover
temperature of the RPM is similar to that of ordinary fluids. 

Although the study of criticality in the RPM has proceeded for half a
decade 
\cite{Fisher:94,Fisher-Levin:93,Carvalho-Evans:95,Fisher-Lee:96,Stell:92,Stell:95,Guillot:94,Fisher:96,Zuckerman:97,Brilliantov:98,Panagiotopoulos,Caillol-Levesque-Weis:96,Valleu-Torrie:98}
progress has been slow and no definitive 
conclusions have yet been reached.
Fisher and Levin \cite{Fisher-Levin:93} extended the original
Debye-H{\"u}ckel theory and established the importance of including dipolar 
Bjerrum pairs \cite{Zuckerman:97} and of accounting for the solvation of
the
neutral, but
electrostatically active, pairs in the screening ionic fluid. These
authors have obtained the best theoretical estimates of the critical
density and temperature of the
RPM when
compared with the results of computer simulations.
\cite{Fisher:96,Panagiotopoulos,Caillol-Levesque-Weis:96,Valleu-Torrie:98}

However, the direct Fisher-Levin 
approach does not address the question of the universality class of the
RPM. On the other hand, computer simulations of the critical behavior of the 
RPM have
been carried out; but these also have not provided clear-cut answers.
Calliol {\it et al.} \cite{Caillol-Levesque-Weis:96} 
have obtained evidence of consistency with Ising exponents, as have
Orkoulas and Panagiotopoulos; \cite{Fisher:96,Panagiotopoulos} but
Valleau and Torrie \cite{Valleu-Torrie:98} 
claim that no sign of Ising behavior is observed in their simulations
above criticality. 

Nevertheless, it is instructive to pose a more modest question, namely:
``How does the
introduction of weak electrostatic interactions affect the near-critical
behavior of a fluid that, when uncharged, displays Ising criticality ?''
While one cannot discard the possibility that the long-range Coulomb
interactions discontinuously change the universality class,
\cite{Fisher:94} the Debye screening of the ions at any non-zero
ionic strength can be regarded as giving rise to an effective short-ranged
interaction, \cite{Fisher:94,Stell:92,Stell:95} and thus --- at least in a
perturbative sense --- we may suppose that weak electrostatic couplings
will not change the character of the transition. They must, nonetheless,
affect all the {\em non-universal} critical properties: and many of these
are of
independent interest.

Accordingly, in this article, we address this last issue. Specifically, we
consider an ordinary fluid of dielectric constant $D$ with only
short-range, or van der Waals interactions, that exhibits criticality of
Ising character at a temperature $T_c^0$. For simplicity in this
essentially exploratory study, we focus on a fully symmetric model in
which the particles of the original, uncharged fluid are all identical
having repulsive cores of diameter $a$ and fixed attractive pair
interactions, the strength of which is measured by $k_B \, T_c^0$. 

Then we ask how the reduced crossover temperature, $t_\times$, changes
when electrostatic interactions are switched on so that half the particles
become positive ions of charge $+e$ and the rest negative ions carrying
charge $-e$. The strength of the Coulomb forces may be measured by the
``contact energy'' $e^2/D \, a$, so that one can imagine either increasing
$e$ from zero or, for fixed charges, decreasing the dielectric constant
$D$ from very high values. Naturally, the critical temperature of the
system will change and it is thus reasonable to define the {\em ionicity}
of the critical charged system by
\begin{equation}
\label{ionicity}
{\cal I} \equiv \frac {e^2}{D \, a \, k_B \, T_c}= \frac {1}{T^*_c},
\end{equation}
where $T^*=D \, a \, k_B \, T/e^2$ is the standard definition of reduced
temperature
in primitive model electrolytes. \cite{LeveltSengers-Given:93,Fisher:94}
The ionicity will constitute our perturbative parameter.

To treat the model, we first consider, in Section II, the purely ionic
limit (where $k_B \, T_c^0=0$) and, using a lattice cutoff, reformulate
the
Coulomb interactions via a quadratically truncated sine-Gordon
transformation. \cite{Glimm-Jaffe:87,remark_1} This corresponds to a
Debye-H{\"u}ckel limiting-law level of analysis although the expansion
could be carried to higher order. \cite{remark_1} Then, in
Section III, we introduce (charge independent) short-range interactions
with the aid of a Kac-Hubbard-Stratonovich (KHS)
\cite{Fisher:83,Binney:92} transformation. In the resulting (approximate)
Hamiltonian we integrate out the sine-Gordon charge field $\psi(\bf r)$,
perturbatively to obtain an effective Landau-Ginzburg-Wilson (LGW)
Hamiltonian in terms of a spin (or density) field, $s(\bf r)$, but with
coefficients depending on the ionicity, $\cal I$. Finally, effectively
truncating at the $s^4$ level, we estimate the crossover temperature 
with the aid of a Ginzburg criterion 
\cite{Carvalho-Evans:95,Fisher-Lee:96,Goldenfeld:92} in Section IV.
This treatment
indicates that, in general, the reduced crossover temperature decreases by 
an amount of order $25\%$. Such a weak change is far too small to
describe the experimental observations although the overall trends seem 
to correlate: see Table II and the discussion and conclusions in
Section V. 


\section{lattice cutoff model for the ionic fluid}
\label{model}

Consider a system of unit-valence positive and negative ions with
repulsive cores but otherwise interacting only through Coulomb
forces. To account approximately for the ionic cores we may suppose the
ions reside on the sites of a lattice of spacing $a$ equal to the atomic
diameter. Alternatively, in momentum space, we can employ a cutoff on a
corresponding Brillouin zone or, for convenience, approximate this by a
sphere of radius $\pi/a$. The grand partition function for the lattice
system with sites labelled $i$, $j$, etc., is 
\begin{equation}
  \label{partition_1}
  \begin{split}
    {\mathcal Z} =& \sum_{n_+,n_-} \exp \Bigl\{-\mu_+ \, n_+ -\mu_- \, n_- \Bigr\}
    \times \\ &\sum_{\{\rho_+,\rho_-\}}
    \exp \Bigl\{-\frac{1}{2} \, \sum_{i,j} [\rho_+(i)-\rho_-(i)] \,
    \varphi_c(i,j) \, [\rho_+(j)-\rho_-(j)] \Bigr\},
  \end{split}
\end{equation}
where the first sum runs over $n_+$ and $n_-$, the total numbers of
positive and negative ions, subject to $n_+ + n_- \leq N$, the
number of lattice sites, while the second sum runs over different 
lattice configurations, for  given numbers, $n_+$ and $n_-$. The
occupation variable $\rho_+(i)$, is 1 if a positive ion is at site $i$,
but
0 otherwise, with a similar definition of $\rho_-$.
We omit the possibility of ionic pairing (molecule formation)
\cite{Fisher:94} so that a lattice site cannot be occupied by more than a
single ion.
The remaining symbols have their
usual meanings: $\mu_+$ and $\mu_-$ being the chemical potentials (divided
by
$k_B \,T$) for the positive and negative ions, respectively.

Since this is a first approach and we are interested in real fluids
(rather than details induced by the lattice
structure) we do not take the Coulomb interaction proportional
to the lattice Green's function but simply use the reduced
three-dimensional isotropic Coulomb potential so that in 
(\ref{partition_1}) we have
\begin{equation}
  \label{coul}
  \varphi_c(i,j)= \frac{e^2} {k_B \, T \, D \, r_{ij} },
\end{equation}
where $r_{ij}$ is the distance separating sites $i$ and $j$.

Given a positive definite matrix $A_{ij}$, the identity
\begin{equation}
  \label{KHS_2}
  \exp \Bigl\{ - \frac{1}{2} \sum_{i,j} y_i \, A_{ij} \, y_j \Bigr\} =
  \sqrt{\frac{\det [ A_{ij} ]}{(2 \pi)^N}} \int\limits_{-\infty}^{+\infty}
  \Bigl( \prod_i dx_i \Bigr) \; 
  \exp \Bigl\{ - \frac{1}{2} \sum_{i,j} x_i \, A_{ij}^{-1} \, x_j - 
  {\mathtt i} \, \sum_{i} x_i \, y_i \Bigr\},
\end{equation}
(where $\mathtt i = \sqrt{-1}$) may be used to rewrite the grand partition
function of a repulsive system with discrete variables in terms of the
partition function of a system with continuous variables: this constitutes
the well known sine-Gordon transformation \cite{Glimm-Jaffe:87} that may
be regarded as 
a modified form of the Kac-Hubbard-Stratonovich transformation used
for attractive interactions. \cite{Fisher:83,Binney:92} These
transformations are standard and need not be
discussed in further detail here.

The diagonal elements of the electrostatic interaction
matrix, $\varphi_c(i,i)$, are chosen to guarantee that $\varphi_c(i,j)$
is positive definite, \cite{Glimm-Jaffe:87} but are otherwise arbitrary 
since a lattice site cannot be occupied by more than a single ion. Thus  
(\ref{KHS_2}) may be used to write (\ref{partition_1}) as, 
\begin{equation}
  \label{partition_2}
  {\mathcal Z} = \int \frac{D\psi}{Z_0} \, 
  \exp \Bigl\{-\frac{1}{2} \, \sum_{i,j} \psi(i) \,
    \varphi_c^{-1}(i,j) \, \psi(j) \Bigr\} \, {\mathcal W}_0,
\end{equation}
where the weight function is given by
\begin{equation}
  \label{w_func_1}
  {\mathcal W}_0 = \sum_{n_+,n_-} \sum_{\{\rho_+,\rho_-\}} 
  \exp \Bigl\{-{\mathtt i} \sum_{j} [\rho_+(j)-\rho_-(j)] \, \psi(j)
  -\mu_+ \, n_+ -\mu_- \, n_- \Bigr\}.
\end{equation}
The continuous field $\psi(j)$, ranging from $-\infty$ to
$+\infty$, is conjugate to the (reduced) discrete charge
density
\begin{equation}
\label{charge}
 q(j)=\rho_+(j) - \rho_-(j).
\end{equation}
On the lattice one has $D\psi=\prod_j d\psi(j)$, while $Z_0$ is an
unimportant constant given by  
\begin{equation}
  Z_0 = \int D\psi \, 
  \exp \Bigl\{-\frac{ 1}{2} \, \sum_{i,j} \psi(i) \,
    \varphi_c^{-1}(i,j) \, \psi(j) \Bigr\} = 
    \sqrt{\frac{ (2 \, \pi)^N}{\det[\varphi_c(i,j)]}}.
\end{equation}

Defining the thermodynamic fields $\mu \equiv (\mu_+ + \mu_-)/2$
and $\gamma \equiv (\mu_+ - \mu_-)/2$, and noting that the total ionic
density may be written as 
\begin{equation}
\label{density}
\rho(j)=\rho_+(j) + \rho_ -(j)=q^2(j), 
\end{equation}
the weight function takes the more convenient form
\begin{equation}
  \label{w_func_2}
  {\mathcal W}_0 = \sum_{q=-1,0,1} \,   
  \exp \Bigl\{-{\mathtt i} \sum_{j} q(j) \, 
  [\psi(j) + {\mathtt i} \, \gamma] - \sum_{j} q^2(j) \, 
  \mu\Bigr\}.
\end{equation}
Carrying out the sum over $q$ here, leads to
\begin{equation}
  \label{w_func_3}
  \begin{split}
    {\mathcal W}_0 &= \prod_j \Bigl( 1 + \exp\{-{\mathtt i}
[\psi(j)+{\mathtt i} 
    \gamma] - \mu \} + \exp\{+{\mathtt i} [\psi(j)+{\mathtt i} 
    \gamma] - \mu \} \Bigr) \\
    &= \exp \Bigl\{ \sum_j \ln \bigl[ 1 + 2 \, \exp(-\mu ) \, 
    \cos(\psi(j)+{\mathtt i} \gamma ) \bigr] \Bigr\}.
  \end{split}
\end{equation}

It may now be checked, using (\ref{partition_2}) and (\ref{w_func_3}), 
that global charge neutrality, namely $\langle q \rangle = 0 $, is
obtained if
$\gamma = 0$, i.e., when the chemical potentials of the positive and
negative
ions are the same. From hereon we assume this to be so and set
$\gamma=0$.

Provided $z \equiv e^{-\mu} < \frac{1}{2}$, we may expand the logarithm in
the weight function ${\mathcal W}_0$ with respect to $\psi(j)$. By
truncating the expansion at second order \cite{Fisher:83} we find
\begin{equation}
\bigl({\cal Z}/{\cal Z}_{LL}\bigr)^{1/N} \approx 1+2\,z+\cdots,
\end{equation}
where we have defined the Debye-H{\"u}ckel limiting-law partition function
as, 
\begin{equation}
  \label{partition_3}
  \begin{split}
    {\mathcal Z}_{LL} &= \int \frac{D\psi}{Z_0} \, 
    \exp \Bigl\{-\frac{1}{2} \, \sum_{i,j} \psi(i) \,
    \bigl[ \varphi_c^{-1}(i,j) + \frac{2 \, z}{(1+2 \, z)} \delta(i,j)
    \bigr] \, \psi(j) \Bigr\} \\
    &= \int \frac{D\psi}{Z_0} \, 
    \exp \Bigl\{-\frac{1}{2} \, \sum_{i,j} \psi(i) \,
    \varphi_{LL}^{-1}(i,j) \, \psi(j) \Bigr\},
  \end{split}
\end{equation}
where $\delta(i,j)$ is the Kronecker delta and 
$\varphi_{LL}(i,j)$ is the Debye-H{\"u}ckel limiting-law (DHLL)
reduced effective potential as we will demonstrate.

Now, Fourier transformation of the electrostatic potential (\ref{coul}) on
the lattice leads to
\begin{equation}
  \label{elec_k}
\hat  \varphi_c(k) = \frac{4 \, \pi  \, \ell_B}{k^2} [1 + O (k^2 a^2)],
\end{equation}
where $\ell_B = e^2 / D \, k_B \, T = a / T^*$ is the Bjerrum length
(the
distance
at which two elementary charges  interact with an energy $k_B \, T$). 
The $O(k^2 a^2)$ terms here arise from the lattice structure or,
equivalently within our approximation, from the repulsive core of the
ion-ion interactions. Since our subsequent analysis will utilize only the
leading $k \rightarrow 0$ behavior, the effects of the lattice structure
entering this way, {\em via} the Coulombic potential, should not be
significant; thus we will neglect the $O(k^2 a^2)$ term in
(\ref{elec_k}). Alternatively, one may regard the pure $1/k^2$ behavior
to be retained in (\ref{elec_k}) as defining (via inverse Fourier
transformation) the Coulomb potential as viewed on the lattice in real
space: this would mean that $\varphi_c(i,j)$ as stated in (\ref{coul})
should be modified slightly on distance scales of order $a$. However, in a
fuller, more precise treatment, of the lattice primitive model it is
possible that such contributions could play a more crucial
role;
\cite{Ciach-Stell:98} but that issue lies outside the scope of our present
analysis.

It follows from (\ref{partition_3}) and (\ref{elec_k}) that the DHLL
effective potential in Fourier space is given by
\begin{equation}
  \label{DH_pot}
  \frac{\hat \varphi_{LL}(k)}{4 \, \pi \, \ell_B(T)} = 
  \frac{1}{k^2+ 4 \, \pi \, \ell_B \, \eta},
\end{equation}
where $\eta = 2 \, z / (1+ 2 \, z)$ with $z \equiv e^{-\mu}$ and, for
convenience here and in most places below, we express $k$, $\ell_B$, etc.,
using length units for which $a \equiv 1$. By differentiating
the grand partition function (\ref{partition_2}) with respect to $\mu$ and
expanding the result in powers of $\psi$, we find 
the total density of ions is given by 
\begin{equation}
\rho \equiv \langle q^2(j) \rangle =
  \frac{2 \, z}{1 + 2 \, z} -  
  \frac{z \, (1+2 \, z+2 \,z^2)}{(1+z) \, (1+2 \, z)^2} \, \langle
\psi_j^2
\rangle + \cdots.
\end{equation}
Retaining only the first term in this series, which implies $\eta
\approx \rho$, and transforming 
(\ref{DH_pot}) back into real space, we obtain
\begin{equation}
  \label{DH_pot2}
  \varphi_{LL}(r) \propto \frac{ \exp (- r / \xi_D ) }{r},
\end{equation}
where $\xi_D = 1/ \sqrt{4 \, \pi \, \ell_B \, \rho}$, is the standard 
Debye screening length, \cite{Debye-Hueckel:23} thus demonstrating that
the correct DHLL effective potential for ionic systems may be obtained from a 
quadratic expansion of the sine-Gordon weight function, in $\psi$ --- a
well known result. \cite{Glimm-Jaffe:87,remark_1}

In the next section we derive, using similar techniques,
a field-theoretic description of a system with additional short-range
forces. As explained, we will focus on the situation where these 
forces drive the critical behavior. Then by integrating out the
electrostatic interactions at the DHLL level we obtain 
corrections to the nonuniversal critical parameters, including an
estimate of the crossover temperature, $t_\times$.


\section{Symmetric Ionic Model with short-range attractions}


By adding to the lattice-cutoff model an 
{\em attractive} symmetric short-range pairwise potential identical for
all ionic pairs, so that 
$\varphi_{++}^s(i,j)=\varphi_{--}^s(i,j)=\varphi_{+-}^s(i,j)=\varphi_s(i,j)$,
the grand partition function becomes 

\begin{equation}
  \label{new_partition_1}
  \begin{split}
    {\mathcal Z} &= \sum_{ \{q\}} \exp \Bigl\{ - \sum_{i}q^2(i)\, \mu 
-\frac{1}{2} \, \sum_{i,j} \Bigl[ q(i) \,
    \varphi_c(i,j) \, q(j) + q^2(i) \,
    \varphi_s(i,j) \, q^2(j)\Bigr] \Bigr\},
  \end{split}
\end{equation}
where we have set $\gamma \equiv (\mu_+ - \mu_-)/2 =0$, to ensure global
charge neutrality. The sum on $\{q\}$ runs over all possible 
configurations of ions on the lattice.

The sine-Gordon transformation may be used as before on the
electrostatic interactions in (\ref{new_partition_1}) to obtain a
description in terms of the field, $\psi$.
In general, the short-range interaction matrix has positive and
negative eigenvalues \cite{remark_2} and thus the KHS
transformation cannot be applied directly. \cite{Fisher:83} Here, we
assume that $-\varphi_s(i,j)$ is positive definite and use the identity
\begin{equation}
  \label{KHS_1}
  \exp \Bigl\{ \frac{1}{2} \,\sum_{i,j} y_i \, A_{ij} \, y_j \Bigr\} = 
  \sqrt{\frac{\det [ A_{ij} ]}{(2 \pi)^N}} \int\limits_{-\infty}^{+\infty}
  \Bigl( \prod_i dx_i \Bigr) \; 
  \exp \Bigl\{ - \frac{1}{2} \sum_{i,j} x_i \, A_{ij}^{-1} \, x_j - 
    \sum_{i} x_i \, y_i \Bigr\}
\end{equation}
to transform the short-range attractive interactions. After these
transformations the grand partition
function (\ref{new_partition_1}) reads
\begin{equation}
  \label{new_partition_2}
  {\mathcal Z} = \int \frac{D\psi \, Ds}{Z_1} \, 
  \exp \Bigl\{-\frac{1}{2} \, \sum_{i,j} \Bigl[ \psi(i) \,
    \varphi_c^{-1}(i,j) \, \psi(j) + s(i) \,
    \varphi_s^{-1}(i,j) \, s(j)\Bigr] \Bigr\} \, {\mathcal W}_1,
\end{equation}
where the field $s(j)$ is conjugate to the total ionic density
$q^2(j)$: see (\ref{density}). The normalizing factor is now
\begin{equation}
  Z_1 =     
   \frac{ (2\, \pi)^N}{\sqrt{\det[\varphi_c(i,j)]\det[\varphi_s(i,j)]}},
\end{equation}
and the new weight function is 
\begin{equation}
  \label{w_func_s}
  {\mathcal W}_1 = \exp \Bigl\{ \sum_j \ln \bigl[ 1 + 2 \, \exp[- s(j) -
\mu] \, 
  \cos\psi(j) \bigr] \Bigr\}.
\end{equation}

Now let us shift (and rescale) the $s$ field by putting
\begin{equation}
  \tilde{s}_j = \tfrac{1}{2} (s_j + \mu - \ln 2),
\end{equation}
and expand the logarithm in (\ref{w_func_s}) in powers of $\psi$ and
$\tilde s$. For $\psi=0$ this shift 
eliminates the odd powers of $\tilde s$ (of order higher than unity) from
the expansion, thereby simplifying subsequent calculations.
We will truncate the expansion at second order in $\psi$, thus
treating the electrostatic interactions only at the DH limiting-law level.
The
expansion in $\tilde{s}$ will be retained to fourth order: we
assume $\tilde s (\bf r)$ to be the basic order-parameter field and, as in
the usual LGW formulation, \cite{Fisher:83,Binney:92}  this is the minimum
number of terms required to study simple critical points. 
The grand partition function then reads
\begin{equation}
  \label{new_partition_3}
  \begin{split}
    {\mathcal Z} & \approx \int \frac{D\psi \, Ds}{Z_1} \, 
    \exp \Bigl\{-\frac{1}{2} \, \sum_{i,j} \psi(i) \,
    \Bigl[ \varphi_c^{-1}(i,j) + \frac{\delta(i,j)}{2} \Bigr] 
    \, \psi(j)  + \frac{1}{4} \, \sum_j (\tilde{s}_j -
\tfrac{1}{3}\tilde{s}_j^3) \, \psi_j^2 \\
    & \, \, \, \, \, \, \, \, \, \, \, \, \, \, \, \, \, \, \, \, \,  -
{\tilde \mu}\sum_j{\tilde s} (j) -\frac{1}{2} \, \sum_{i,j} \tilde{s}(i)\,
    \Bigl[ 4\, \varphi_s^{-1}(i,j) - \delta(i,j)\Bigr] \, \tilde{s}(j) - 
    \frac{1}{12} \sum_j \tilde{s}_j^4 \Bigr\},
  \end{split}
\end{equation}
where we have put ${\tilde \mu}=1+2 \, (\ln 2 -\mu)
\sum_j \varphi_s^{-1}(0,j)$.

In order to perform the integration over the
non-critical field $\psi$, we rewrite the partition function in reciprocal 
space, replacing lattice sums by integrals, using 
\begin{equation}
  \frac{1}{N} \sum_k \approx \int\limits_{\Omega} \frac{d^3k}{(2 \,
\pi)^3}
  \equiv \int\limits_k \, \, \, \, \, \, \, \, \, (N \rightarrow
\infty),
\end{equation}
where $\Omega$ is the first Brillouin zone of the reciprocal lattice.
\cite{Fisher:83,Binney:92}
We further suppose that the Fourier transformed short-range
potential is of the form \cite{Fisher:83}
\begin{equation}
\label{sr_potential}
   \hat \varphi_s(k) \approx  4\,J_0\,(1 - R_0^2 \, k^2) ,
\end{equation}
where $J_0$ is related to the strength of the interaction (divided by
$k_B \, T$) and $R_0$ to the interaction range, both of which can be
determined for a given model.
The values of $J_0$ and $R_0$ affect the nonuniversal critical parameters
(temperature, chemical potential and crossover temperature)
but, as we will see below, estimates of the crossover temperature, 
$t_\times$, may be obtained by examining various typical values for $R_0$.
As in the discussion following (\ref{elec_k}), we have neglected in
(\ref{sr_potential}) higher order lattice terms, etc., of the form $k^4
a^4$, etc.

Finally, in terms of the ionicity $\cal I$, defined in (\ref{ionicity}) it
is convenient 
to introduce the small perturbation parameter $y$ and to rescale the
$\psi$ field in reciprocal space according to
\begin{equation}
  y^2= 4 \, \pi \, {\cal I} \equiv \frac {4 \, \pi}{T^*}, 
\,\,\,\,\,\,\,\,\,\,\,\, \tilde{\psi}(k) = \frac{\hat \psi(k)}{y}. 
\end{equation}
[Since we are interested only in the critical region, we may, here,
neglect the difference between $T^*$ and $T_c^*$, as used in
(\ref{ionicity}).]
The reduced Hamiltonian, $-H/k_B\,T$, in (\ref{new_partition_3})
then becomes 
\begin{equation}
  \label{action_1}
  \begin{split}
    {\mathcal H} =& -\frac{1}{2} \, \int\limits_k 
    \bigl( \frac{y^2}{2}+ k^2 \bigr) \, |\tilde{\psi}(k)|^2
+ \frac{ y^2}{4} \iint\limits_{k_1\,k_2} \hat s(k_1) \, 
    \tilde{\psi}(k_2) \, \tilde{\psi}(-k_1-k_2)\\  
&- \frac{y^2}{12} \iiiint\limits_{k_1\,k_2\,k_3\,k_4} 
   \hat  s(k_1) \, \hat s(k_2) \, \hat s(k_3) \, \tilde \psi(k_4) \, 
\tilde \psi(-k_1-k_2-k_3-k_4)
- {\Delta \mu_{sr}} \hat s(0)\\
 &-\frac{1}{2} \, \int\limits_k     \bigl(r_{sr}+  \tau^2_{sr}\, k^2
\bigr)\,|\hat s(k)|^2 
- u_{sr} \iiint\limits_{k_1\,k_2\,k_3} 
   \hat  s(k_1) \, \hat s(k_2) \, \hat s(k_3) \, \hat s(-k_1-k_2-k_3),
  \end{split}
\end{equation}
where $\Delta \mu_{sr}=1+ (\ln 2 -\mu) /2\, J_0$  while
\begin{equation}
\label{sr_coefficients}
r_{sr}= \frac{1} {J_0}-1, \,\,\,\,\,\,\,\,\,\,\,\,\tau^2_{sr}=
\frac{R_0^2}
{J_0}, \,\,\,\,\,\,\,\,\,\,\,\,u_{sr}=\frac{1}{12}, 
\end{equation}
are the coefficients of the reduced Hamiltonian of the short-range
uncharged fluid. 

Now on integrating over $\tilde \psi$ in the partition function
(\ref{new_partition_3}),  
the coefficients of $\hat s^n$ in $\cal H$ become renormalized because the
fields are coupled through the cubic and fifth order terms, $\hat s \, \tilde
\psi^2$ and $\hat s^3 \, \tilde \psi^2$.
This integration over $\tilde \psi$ as a simple Gaussian field could be 
carried out exactly but it involves a nontrivial diagonalization of $\cal
H$ which is not required, since the resulting Hamiltonian is to be
expanded only to fourth order in $\hat s$, for consistency with the
previous truncation. (However, the study of Brilliantov
\cite{Brilliantov:98} is instructive in this respect.) In fact, the simplest 
way to calculate the electrostatic corrections to the coefficients of 
$\hat s^n$, is via perturbation theory using Feynman diagrams.
To this end we write the partial reduced Hamiltonian
\begin{equation}
\begin{split}
  {\mathcal H}_{\psi} =& -\frac{1}{2} \, \int\limits_k 
  \bigl( \frac{y^2}{2} + k^2 \bigr) \, |\tilde{\psi}(k)|^2
  + \frac{y^2}{4}   \iint\limits_{k_1,k_2} \hat s(k_1) \, 
  \tilde{\psi}(k_2) \, \tilde{\psi}(-k_1-k_2)\\
&- \frac{y^2}{12} \iiiint\limits_{k_1\,k_2\,k_3\,k_4} 
   \hat  s(k_1) \, \hat s(k_2) \, \hat s(k_3) \, \tilde \psi(k_4) \, 
\tilde \psi(-k_1-k_2-k_3-k_4),
\end{split}
\end{equation}
where 
the first term is simply Gaussian while we regard the second and third as
the ``interaction''. 

A standard cumulant expansion \cite{Fisher:83,Binney:92} yields the
$n$-point correlation functions
generated by the Gaussian propagator. By carrying the expansion out to
fourth order, we obtain a Hamiltonian for a single fluctuating field in
the form
\begin{equation}
  \label{hamiltoniano0}
  {\mathcal H}_0=- \frac{1}{2} \int\limits_{ k}
  (r_0 +\tau_0^2 \, k^2) \, |\hat s|^2 
  - v_0 \iint\limits_{ k_1\, k_2}   \hat s \, \hat s \, \hat s \,
  - u_0 \iiint\limits_{ k_1\, k_2\, k_3} 
  \hat s \, \hat s \, \hat s \, \hat s - \Delta \mu_0 \, \hat s_0 .
\end{equation}
where the coefficients, renormalized by the electrostatic
interactions, are given to leading orders by
\begin{equation}
  \label{r0}
  r_0 = r_{sr} - \tfrac{1}{8} \, y^4 \, I_2(y), 
\end{equation}
\begin{equation}
  \label{tau0}
  \tau_0^2=\tau_{sr}^2 + \tfrac {1}{8} \, y^4 \, I_{3,4}(y) ,
\end{equation}
\begin{equation}
\label{v_0}
v_0=\tfrac{1}{12} \, y^2 I_1(y) -\tfrac{1}{48} \, y^6 \,I_3 (y),
\end{equation}
\begin{equation}
  \label{u0}
  u_0 = \tfrac{1}{12} \, [1+ \tfrac{1}{2} \, y^4 \, I_2(y) \,
-\tfrac{3}{32}\, y^8 \, I_4(y) ],
\end{equation}
\begin{equation}
  \label{h0}
   \Delta \mu_0 = {\Delta \mu_{sr}} - \tfrac{1}{4} \, y^2 \, I_1(y).
\end{equation}
If the first Brillouin zone of the reciprocal
lattice is approximated by a sphere of radius $\pi/a$ the propagator
integrals, $I_n(y)$, are given by
\begin{equation}
  \label{gama_n}
  I_n \simeq 
  \frac{1}{2 \, \pi^2} \int\limits_{0}^{\pi} dx
  \frac{x^2}{(x^2 + y^2/2)^n} ,
\end{equation}
\begin{equation}
  \label{gama_3,4}
  I_{3,4} \simeq \frac{1}{2 \, \pi^2} \,
  \int\limits_0^{\pi} dx \: \Big\{ \frac{x^2}{(x^2+y^2/2)^3} - 
  \frac{4 \, x^4}{3\, (x^2+y^2/2)^4} \Big\}.
\end{equation}
Clearly, when the ionicity vanishes, $y \rightarrow 0$, the
coefficients $r$, $\tau$, $\cdots$  reduce to those of the
short-range, uncharged fluid.

Finally, the third order term in (\ref{hamiltoniano0}) may be
eliminated in the usual way, $\hat s
\Longrightarrow \hat s + v_0 / 4 \, u_0$, yielding a standard
short-range LGW Hamiltonian, namely,
\begin{equation}
  \label{hamiltoniano}
  {\mathcal H}_s=- \frac{1}{2} \int\limits_{ k}
  (r +\tau^2 \, k^2) \, |\hat s|^2 
  - u \iiint\limits_{ k_1\, k_2\, k_3} 
  \hat s \, \hat s \, \hat s \, \hat s - \Delta \mu \, \hat s_0,
\end{equation}
with coefficients, 
\begin{equation}
  \label{coefficients}
r = r_0 - \frac{3\,v_0^2}{4\, u_0},\,\,\,\,\,\,\,\,\,\,\,\,\tau^2=\tau_{0}^2,
\,\,\,\,\,\,\,\,\,\,\,\, u = u_0,\,\,\,\,\,\,\,\,\,\,\,\,\Delta \mu ={\Delta \mu_0}-\frac{v_0\,r_0}{4
\,u_0}+\frac{v_0^3}{8\,u_0^2}.
\end{equation}

Accepting the limitations imposed and the approximations made, we are now
in a position to study the effects of the electrostatic
interactions on the critical behavior of a simple fluid. The easiest way
to proceed, as indicated above,  is to use the Ginzburg criterion to
estimate the crossover 
temperature, $t_\times$, as a function of the ionicity ${\cal I} \propto
y^2$. 
This criterion indicates the size of the temperature interval around the
critical point outside which a mean-field description is
self-consistent.
For temperatures inside that interval the mean-field description
breaks down and Ising behavior (in this case) should manifest itself. 
The temperature scale so set has become known as the
Ginzburg temperature, $t_G$, and it is reasonable to take it as an
estimate of the
crossover temperature, $t_\times$,
\cite{Fisher-Levin:93,Carvalho-Evans:95,Fisher-Lee:96} although the
presence of significant higher-order terms in (\ref{action_1}) could
undermine the practical validity of the criterion.
\cite{Fisher-Lee:96,Brilliantov:98} 
In the next section we compare the Ginzburg temperatures of
systems with finite ionicity (${\cal I} \propto y^2 > 0$) and without 
electrostatic interactions ($y=0$).


\section{Application of the Ginzburg Criterion}


The condition for a self-consistent mean-field 
description is that fluctuations of the field (calculated using the mean
field) are much smaller than the typical or average mean field.
For the $s^4$
Hamiltonian (\ref{hamiltoniano}) studied below $T_c$, 
this condition may be expressed conveniently as \cite{Goldenfeld:92}  
\begin{equation}
  \label{ginz}
  \frac{3 \, \sqrt{2}}{\pi} \, \frac{u}{|r|^{1/2} \, \tau^3} \ll 1 .
\end{equation}
Naturally, in deriving this, we have set $\Delta \mu=0$, so that the
critical point is approached along an axis of asymptotic symmetry. It is then
appropriate to note that the short-range coefficient, $r_{sr}$, can
normally be written as \cite{Fisher:83}
\begin{equation}
  \label{r_sr}
  r_{sr} = \frac{T-T_{c,0}^0}{T_{c,0}^0} ,
\end{equation}
where $T_{c,0}^0$ is the mean-field critical
temperature of the uncharged system. 
For a system with electrostatic interactions the mean-field critical 
temperature, $T_{c,0}(y)$, increases by an amount which follows from 
(\ref{r0}) and (\ref{coefficients}) as
\begin{equation}
  \label{t_0}
t_0(y)\equiv \frac{T_{c,0}(y)}{T_{c,0}^0} -1= \tfrac{1}{8} \, y^4 \, I_2(y)
+ \frac{3\,v_0^2}{4\, u_0}.
\end{equation}
We may use this result to rewrite the coefficient $r$ in (\ref{ginz}) as
\begin{equation}
  r = \frac{T-T_{c,0}(y)}{T_{c,0}^0}=[1 + t_0(y)] t,
\end{equation}
where $t=(T-T_{c,0})/T_{c,0}$ is the (mean-field) reduced temperature
of the charged system.
The Ginzburg criterion then becomes
\begin{equation}
  \label{ginz2}
  t_G({\cal I}) \simeq \frac{18}{\pi^2} \, \frac{u^2(y)}{[1+ t_0(y)]
\tau^6(y)} \ll |t|,
\end{equation}
which defines the reduced Ginzburg temperature, $t_G$. At this
point the value of $t_G$ depends on the ionicity ${\cal I} = y^2 /4 \pi$
through $\tau(y)$, $u(y)$ and $t_0(y)$, as given in
(\ref{tau0}), (\ref{u0}), (\ref{coefficients}) and
(\ref{t_0}), respectively, and on the range, $R_0$ through
$\tau_{sr}$. The dependence of $t_G$ on $J_0$ has dropped out since, using
(\ref{sr_coefficients}), one finds $\tau_{sr}^2 = [1+t_0(y)]\,R_0^2$. 
 
For the short-range uncharged system the Ginzburg temperature reduces
to
\begin{equation}
t_G(0)=\frac{1}{8\pi^2} \frac{a^6}{R_0^6}.
 \end{equation}
If, as a convenient and not unreasonable value, one adopts $R_0/a \simeq
0.4828$, one obtains $t_G(0)=1$. 
For this particular short-range model, we have calculated the
parameters of the reduced Hamiltionian (\ref{hamiltoniano}) of the system
with electrostatic interactions, as a function of the 
ionicity, ${\cal I}$. The results for $\tau$, $u$ and the shift in the
mean field critical temperature of the system are plotted in Fig.~1, as
functions of ${\cal I}$. 
The corresponding Ginzburg temperature, $t_G({\cal I})$, is displayed in 
Fig.~2. 
The effective interaction range, $\tau$ increases monotonically with
ionicity while the coefficient, $u$, of the fourth order
term exhibits a maximum at ${\cal I} \simeq 3$ and decreases slowly at
higher ionicities. However, 
within our Debye-H{\"u}ckel limiting-law
approximation, $u$ remains positive for all ionicities  with, in fact, $u
\rightarrow \frac{1}{12}(1 + \frac{1}{12} \pi) \simeq 0.105$ as $y
\rightarrow \infty$. 
As a result the Ginzburg temperature, $t_G$, varies nonmonotonically with 
ionicity, ${\cal I}$: see Fig.~2. A maximum, $t_G \simeq 1.03$, occurs at
${\cal I} \simeq 0.7$ indicating that, in general, the electrostatic
interactions reduce (by up to $\simeq 30\%$) the interval around $T_c$ in
which
Ising critical exponents may be observed. Closer inspection of the
behavior of $t_G$ with $\cal I$ reveals an initial drop off of $t_G$, as
shown in the inset. 

In order to check the robustness of these results, we have calculated the
ratio of reduced Ginzburg temperatures, $t_G(y)/t_G(0)$, for other values
of $R_0/a$ in the range 0.1 to 10 (corresponding to uncharged Ginzburg
temperatures, $t_G(0)$, spanning 12 orders of magnitude). The results are
plotted in Fig.~3. The behavior of the ratio of Ginzburg temperatures,
$t_G(y)/t_G(0)$, varies with the particular value of $R_0$ but at
ionicities ${\cal I} \ge 2$, a reduction of $t_G$ is found for 
all the systems.
At low ionicities, however, the behavior of $t_G$ depends on the range of
the nonionic interactions. The initial drop off is more rapid for low
values of $R_0/a$. For the system with $R_0/a =0.1$ a minimum value of
$t_G$ is clearly visible. 

In assessing the results shown in Figs. 1--3, the perturbative nature of
our calculations must be borne in mind. The scale of probable validity in
${\cal I}$ may be guaged by noting that Monte Carlo results for the RPM,
\cite{Fisher:96,Panagiotopoulos,Caillol-Levesque-Weis:96,Valleu-Torrie:98} 
the extreme limit of ``complete ionicity,'' indicate ${\cal I} \simeq 
20.4$. Clearly our results are, at best, of qualitative validity
when ${\cal I}$ exceeds, say, 5.

Finally, some contact with experimental results may be made by estimating
the values 
of ${\cal I}$ using the observed reduced critical temperatures and other data: see  
Table~\ref{tab:exper} which presents the Ginzburg temperatures calculated for 
various experimental systems (assuming, for simplicity, $R_0/a=0.4828$) and the 
corresponding, experimentally assessed \cite{Pitzer:private} crossover 
temperatures, $t_\times$. 
Realistic values of ${\cal I}$ lead to the prediction of reductions of
the reduced Ginzburg temperature that correlate with the experimental
data. However, our theoretical results for $t_G$ grossly
underestimate the reductions in $t_\times$ observed in the real ionic
systems listed.


\section{Discussion and conclusions}


Starting with a Hamiltonian with attractive short-range as well as
long-range  Coulomb
interactions, we have derived a field-theoretic description for a
symmetric 
model
with two coupled fluctuating fields, $\psi$ (conjugate
to the charge density) and $s$ (conjugate to the number density). 
This lays a foundation on which to study explicitly the influence on
critical properties of the {\em ionicity}, that is the strength of the
Coulomb coupling relative to the short-range interactions. In an
exploratory approach we expanded the weight function to second order in
$\psi$, corresponding to a Debye-H{\"u}ckel limiting-law level, and to
fourth order in $s$, the usual LGW level. Finally, we 
integrated out the $\psi$ field so generating electrostatic
corrections to
the coefficients of $s^n$ in the final effective Hamiltonian, ${\cal
H}_s$ --- see (\ref{hamiltoniano}) --- in a systematic way. 

We examined the perturbative effects of the electrostatic
interactions on the near-critical behavior of a simple fluid by using the
Ginzburg
criterion to estimate the crossover 
temperature, $t_\times$, as a function of the ionicity, defined more
explicitly by ${\cal I}=e^2/D \, 
a \, k_B \, T$: see (\ref{ionicity}). The calculations indicate that as
${\cal I}$ increases, the effective range of 
interaction, $\tau$, increases slowly while the coupling of the fourth
order term in ${\cal H}_s$, namely $u$, decreases for 
realistic values of $\cal I$ and $R_0$ after an initial, relatively rapid
transient. 
As a result, the relative Ginzburg scale, $t_G({\cal I})/t_G(0)$, [see
(\ref{ginz2})] decreases with increasing ${\cal I}$, by up to $20-30\%$ 
(within the approximations): see Fig.~3. This suggests 
that the expected domain of Ising-type behavior around $T_c$
should, likewise, be reduced by the presence of Coulombic forces.

Although we believe these results are interesting, one must certainly be
cautious in applying them to real ionic systems. 
First, it is not certain that the Ginzburg criterion provides a
sufficiently reliable measure of 
$t_\times$.\cite{Singh-Pitzer:90,Fisher:94,Zhang-al:92,Wiegand-al:97,Japas-LeveltSengers:90,Narayanan-Pitzer:94,Pitzer:private} 
Indeed, despite the correct
trend of $t_G$ with the ionicity $\cal I$ seen in Table II, the predicted magnitudes of the 
reductions of $t_\times$ are vastly smaller than observed in the real systems studied.

Second, in our perturbative treatment 
of the model the critical behavior is always driven by the short-range
interactions, and only nonuniversal critical parameters are affected by
the electrostatic interactions. 
It is
unlikely that such a mechanism describes satisfactorily the criticality of 
all real ionic systems. In fact, as argued by Pitzer, \cite{Pitzer:90}
there are chemical systems in which the ionic interactions are almost
surely the principal determinants of the critical behavior. And, of
course, in the restricted primitive model (RPM) only Coulombic forces act
but a critical point is surely realized. 
\cite{Fisher:94,Fisher-Levin:93,Carvalho-Evans:95,Fisher-Lee:96,Stell:92,Stell:95,Guillot:94,Fisher:96,Zuckerman:97,Brilliantov:98,Panagiotopoulos,Caillol-Levesque-Weis:96,Valleu-Torrie:98}
Such systems clearly lie outside the range of validity of our truncated,
perturbative treatment. Furthermore, as suggested both theoretically
\cite{Fisher:94,Ciach-Stell:98} and by the experiments of Pitzer and
Narayanan, \cite{Narayanan-Pitzer:94} Coulombic electrolytes of high
ionicity may lie on a different global thermodynamic locus than do
solvophobic systems such as we have sought to describe in our present
approach.

Within the approximation presented, $u$ never
vanishes; however, it is possible that carrying the calculations to
higher order \cite{Brilliantov:98} would lead to negative $u$ for
sufficiently large ionicity. At a mean-field level in a simple lattice
system that would suggest tricriticality \cite{Nelson-Fisher:75} and the
presence of an ``antiferromagnetic'' or crystal phase.
But, by any measure, one would then need to retain sixth and, possibly,
higher order terms in the effective Hamiltonian; \cite{Nelson-Fisher:75}
and, in that case, a renormalization group treatment would seem necessary
to elucidate the nature of the resulting transition(s).
\cite{Brilliantov:98,Nelson-Fisher:75} It should be forcibly stressed,
however, \cite{Fisher:94} that a tricritical point {\em per se} cannot
arise in a normal one- or two-component fluid system except by some
special, nongeneric accident! (In a three-component system a tricritical
point can occur but is characterized by a particular temperature, $T_t$,
and three fixed densities, say, $\rho_A^t$, $\rho_B^t$ and $\rho_C^t$.)

The present approach has been extended \cite{Moreira:98} to {\em
asymmetric} models with {\em distinct} short-range interactions,
$\varphi_{++}^s({\bf r})$, $\varphi_{+-}^s({\bf r})$
and $\varphi_{--}^s({\bf r})$. The calculations yield an effective
Hamiltonian very similar to that obtained by Nelson and Fisher
\cite{Nelson-Fisher:75} in their study of metamagnets in the presence of
an external magnetic field (which, when large enough, induces
tricriticality). A renormalization group calculation employing the
$\epsilon$-expansion has been performed: \cite{Moreira:98} it yields an
unstable Gaussian fixed point, a symmetric Ising-type (Fisher-Wilson
$n$=1)
\cite{Fisher:83,Binney:92} fixed point, and two further {\em asymmetric}
Ising fixed points. In addition, a separatrix appears, at the $s^4$ level
of approximation, that delimits a region of ``runaway'' flows: such
behavior requires proper study of the Hamiltonian with higher order terms
retained. \cite{Nelson-Fisher:75} However, as hinted above, the
significance of such flows within the context of fluid systems is unclear
at this stage.


\section*{Acknowledgments}


MMTG and AGM thank Alina Ciach for enlightening discussions and are
grateful to the Funda{\c c}{\~a}o para a Ci{\^e}ncia e a Tecnologia (FCT) 
for partial support through a running grant and project PRAXIS 
XXI/2/2.1/FIS/181/94. 
AGM acknowledges the support of the FCT  
through Grant No. PRAXIS XXI/BM/6898/95. 
MEF is grateful for funding from the U. S. National Science Foundation
(through Grant No. CHE 96-14495) and is indebted to Professor Kenneth S.
Pitzer, alas since deceased, for stimulating correspondence and
discussions. Anatoly B. Kolomeisky and Youngah Park kindly commented on
the draft manuscript. 
  


\begin{table}[t]
  \begin{center}
    \caption{Experimental melting and critical
temperatures, $T_m$ and $T_c$, in various solvents for
      three organic salts: (a) tetra-$n$-pentylammonium bromide; (b) triethyl 
      $n$-hexyl ammonium triethyl $n$-hexyl boride; (c)
tetra-$n$-butylammonium 
picrate.} \bigskip
    \begin{tabular}{cccc}
      \rule{0mm}{5mm}Salt & $T_m$ (K) & Solvent & $T_c$ (K)\\ 
      \tableline
      (a) & $\sim 374.05$ & water\cite{Japas-LeveltSengers:90} & 404.90\\
      (b) & $\sim 251$ & diphenyl ether\cite{Singh-Pitzer:90,Zhang-al:92,%
        Wiegand-al:97} & 309--312\\
      (c) & 364--365 & 1-choloroheptane\cite{Pitzer-deLima-Schreiber:85} & 404.4\\
          &          & 1-tridecanol\cite{Weingaertner-Wiegand-Schroer:92} & 342\\
          &          & 1-dodecanol\cite{Narayanan-Pitzer:94} & $\sim 332$\\
          &          & 1,4-butanediol\cite{Narayanan-Pitzer:94} & $\sim 333$
    \end{tabular}
    \label{tab:Tm_Tc}
  \end{center}
\end{table}

\begin{table}[t]
  \begin{center}
    \caption{Experimentally assessed
      crossover temperatures, $t_\times$, and
      reduced Ginzburg temperatures, $t_G$, calculated assuming an
      uncharged model with $R_0/a = 0.4828$ [or $t_G(0)=1$] 
      for\cite{remark_3}  
      (a) tetra-$n$-butylammonium
      picrate (TBAP) in 1-tridecanol;\cite{Weingartner_1:95}
      (b) TBAP in 1-dodecanol;\cite{Narayanan-Pitzer:94}
      (c) TBAP in 75\% 1-dodecanol plus 25\% 1,4-butanediol;
      \cite{Narayanan-Pitzer:94}
      (d) Na in N$\mathrm{H}_3$;\cite{Weingartner_1:95,Chieux-Sienko:70}
      (e) tetra-$n$-pentylammonium bromide in
      water.\cite{Japas-LeveltSengers:90}} \bigskip
    \begin{tabular}{cccc}
      \rule{0mm}{5mm}System & ionicity, ${\cal I}$ & $t_\times$ & $t_G$ \\ 
      \tableline
      uncharged fluid & 0 & $O(1)$ & $1$ \\
      (a) & 17.9 & $ \sim 10^{-3}$ & $\sim 0.712 $ \\ 
      (b) & 16.8 & $ \sim 0.9 \times 10^{-2}$ & $\sim 0.717$ \\ 
      (c) & 8.9 & $\sim 3 \times 10^{-2}$ & $\sim 0.777$ \\
      (d) &6.97 & $0.6 \times 10^{-2}$ & $\sim 0.807$ \\
      (e) & $\sim 1.4$ & $O(1)$ & $\sim 1$
    \end{tabular}
    \label{tab:exper}
  \end{center}
\end{table}


\begin{figure}
  \caption{(a) The effective interaction range, $\tau$, in the Hamiltonian
    (\ref{hamiltoniano})  as a function
    of the ionicity, ${\cal I}$, for a system with $R_0/a = 0.4828$;
    (b) the fourth order coupling constant, $u$, in (\ref{hamiltoniano})
    as a function of the ionicity; (c)
    the reduced shift of the mean-field critical temperature, $t_0$. Notice 
    that both $u$ and $t_0$ are independent of the value of $R_0/a$.}
\end{figure}

\begin{figure}
  \caption{The reduced Ginzburg temperature $t_G$
    as a function of ionicity, ${\cal I}$, for the short-range model of
    Fig.~1, which is described by $t_G(0)=1$. The inset shows the behavior 
    of $t_G$ close to the origin.}
\end{figure}

\begin{figure}
  \caption{Ratio of the Ginzburg temperatures calculated for a system of
    given ionicity to those for the corresponding short-range models
    $t_G({\cal I})/t_G(0)$ for the cases $R_0/a = 0.1$ (dashed
    line), $= 0.2$ (dashed-dotted line), $= 0.4828$ (full line), 
    and $= 10$ (dotted line).}
\end{figure}

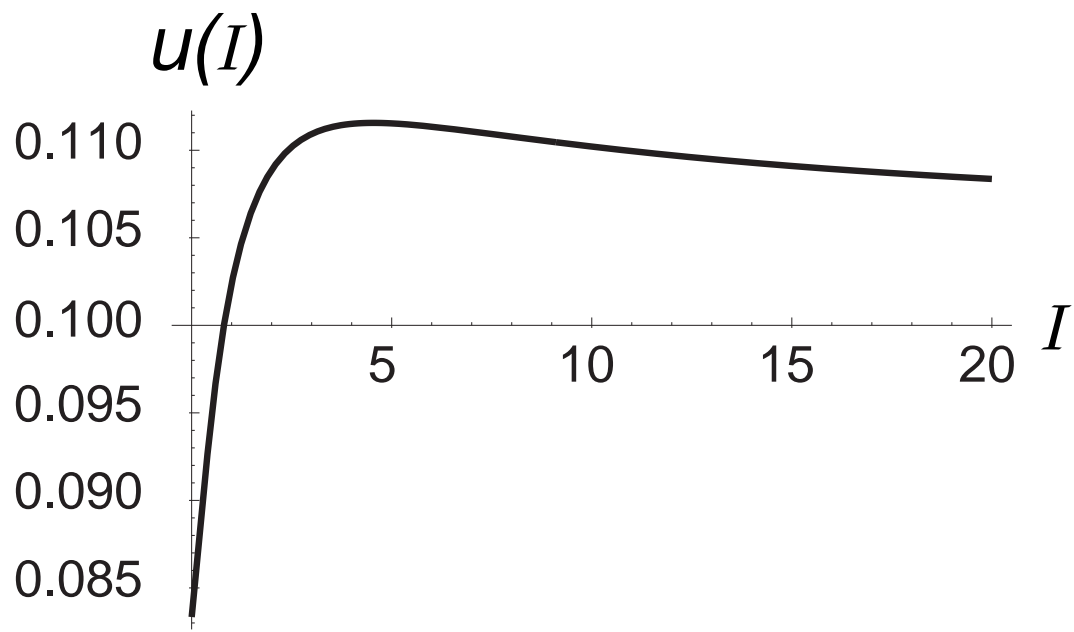
\end{document}